\documentclass[aps,pra,showpacs,superscriptaddress,twocolumn,10pt]{revtex4-1}

\usepackage{hyperref}
\usepackage[cm]{fullpage}
\usepackage{graphicx}
\usepackage[english]{babel}
\usepackage{amsmath}
\usepackage{amssymb}
\usepackage{cancel}
\usepackage{natbib}
\usepackage{comment}
\usepackage{color}
\usepackage[normalem]{ulem}

\newcommand{\ket}[1]{{\left\vert {#1} \right\rangle}}
\newcommand{\bra}[1]{{\left\langle {#1} \right\vert}}

\newcommand{\eq}{Eq.~}
\newcommand{\eqs}{Eqs.~}
\newcommand{\fig}{Fig.~}

\newcommand{\cf} {cf.~}
\newcommand{\ug} {\!=\!}

\newcommand{\piu} {\!+\!}
\newcommand{\meno} {\!-\!}

\newcommand{\eg} {e.g.~}

\newcommand{\rref} {Ref.~}

\newcommand{\Tr}{\text{Tr}}
\newcommand{\U}{\mathcal{U}_\tau}

\newcommand{\compcent}[1]{\vcenter{\hbox{$#1\circ$}}}

\newcommand{\comp}{\mathbin{\mathchoice
  {\compcent\scriptstyle}{\compcent\scriptstyle}
  {\compcent\scriptscriptstyle}{\compcent\scriptscriptstyle}}}

\makeatletter
	\renewcommand{\maketag@@@}[1]{\hbox{\m@th\normalsize\normalfont#1}}
\makeatother

\begin{document}

\author{Salvatore Lorenzo}
\affiliation{Dipartimento di Fisica, Universitˆ della Calabria, 87036 Arcavacata di Rende (CS), Italy}
\affiliation{INFN - Gruppo collegato di Cosenza, Cosenza, Italy}
\author{Francesco Ciccarello}
\affiliation{NEST, Istituto Nanoscienze-CNR and Dipartimento di Fisica e Chimica, Universit$\grave{a}$  degli Studi di Palermo, via Archirafi 36, I-90123 Palermo, Italy}
\author{G. Massimo Palma}
\affiliation{NEST, Istituto Nanoscienze-CNR and Dipartimento di Fisica e Chimica, Universit$\grave{a}$  degli Studi di Palermo, via Archirafi 36, I-90123 Palermo, Italy}

\begin{abstract}
A well-known situation in which a non-Markovian dynamics of an open quantum system $S$ arises is when this is coherently coupled to an auxiliary system $M$ in contact with a Markovian bath. In such cases, while the joint dynamics of $S$-$M$ is Markovian and obeys a standard (bipartite) Lindblad-type master equation (ME), this is in general not true for the reduced dynamics of $S$. {Furthermore, there are several instances (\eg the dissipative Jaynes-Cummings model) in which a {\it closed} ME for the $S$'s state {\it cannot} even be worked out.} Here, we find a class of bipartite Lindblad-type MEs such that the reduced ME of $S$ can be derived exactly and in a closed form for any initial product state of $S$-$M$. We provide a detailed microscopic derivation of our result in terms of a mapping between two collision models.
\end{abstract}


\title{Class of exact memory-kernel master equations}

\maketitle
\section{Introduction}

The ability to manipulate a single, low-dimensional quantum system $S$ is key to the emerging field of quantum technologies. However, due to its unavoidable interaction with the surrounding environment, the dynamics of any realistic quantum system $S$ is {\it open}, i.e. non-unitary, and this is typically detrimental to the effectiveness of such technologies. A thorough and reliable description of the dynamics of an open quantum system \cite{petruccione} is thus of utmost importance, especially in the case of non-Markovian dynamics, as witnessed by the strong current interest in this topic \cite{reviews}.

In the best case, a full description of the reduced quantum dynamics of  $S$ can be given in terms of a closed, well-behaved, master equation (ME) with the density matrix of $S$ as the only unknown. Often, however, this is not the case especially for non-Markovian open dynamics. A paradigmatic and relatively simple example is the decay of an atom in a single-mode lossy cavity as described by the well-known dissipative Jaynes-Cummins model. The bipartite
atom-mode dynamics is governed by a Kossakowski-Lindblad ME \cite{petruccione}, featuring a Hamiltonian term (depending on the atom-mode Hamiltonian) and a Lindbladian dissipator acting on the cavity mode. In this case, tracing out the mode degrees of freedom  {\it does not} lead to a closed ME for the atom \cite{nota-closed}. 
The dissipative Jaynes-Cummins model can be regarded as an instance of a bipartite Lindblad ME in which a quantum system $S$ is coherently coupled to a second one  $M$, the ``memory", the latter interacting with a Markovian bath according to an associated Lindbladian superoperator acting on $M$ only. 

In this paper, we present a class of MEs of the type discussed above which, upon trace over the $M$'s degrees of freedom and for any initial $S$-$M$ product state, yield a closed, exact, ME for $S$. This is an integro-differential memory-kernel ME, defined in terms of the dynamical map of $S$ that would arise if $M$ were decoupled from its environment (hence corresponding to a unitary $S$-$M$ joint dynamics). The resulting memory-kernel ME is well-behaved, meaning that the corresponding dynamics of $S$ is ensured to be completely positive and trace preserving (CPTP) \cite{petruccione}. Very recently, Chruscinski ans Kossakowski \cite{kossa} studied a parametrization of legitimate memory kernels entering the general Nakajima-Zwanzig (NZ) ME \cite{petruccione}, showing that a number of well-behaved non-Markovian MEs can be arranged in the NZ form. We will show that this is true also for our ME.

Furthermore, this ME is not restricted to the scenario in which the system-environment coupling is mediated via the ancillary degrees of freedom $M$ but applies to a broader class of non-Markovian dynamics. Indeed  as we will show, the memory-kernel ME for $S$ discussed in this paper is a  generalisation of a ME first derived through a collision model of non-Markovian open quantum dynamics \cite{ciccarello} (for a different perspective derivation see \rref\cite{vacchini}). A quantum collision model (CM) \cite{CMs} is a microscopic framework to describe the open dynamics of a system $S$ interacting with a reservoir assumed to consist of a large collection of smaller constituents (ancillas). The system is assumed to interact with the environment via a sequence of ``collisions" between system and ancillas, each collision being described by  the same bipartite quantum map (usually a unitary one). The resulting reduced dynamics of $S$ is, by construction, a CPTP map. In the limit of weak coupling this leads to a well-behaved ME. 
CMs have been a useful tool to analyze quantum homogenization and thermalization \cite{scarani}, to derive MEs \cite{buzekME}, to study the interaction with small environments and or with random unitaries \cite{Benenti}, in quantum thermodynamics \cite{thermo} and in the study of quantum non-Markovianity \cite{NMCMs, phys-scrip}. Experimentally, a CM can be implemented in all-optical setups \cite{sciarrino}.
 In the second part of this paper, we will  illustrate two different environmental memory mechanisms  in the context of collision models, both of which  lead to  reduced open dynamics governed by our ME. 
We will first show that -- when appropriately generalised -- the CM without memory $M$ but with inter-ancillary collisions introduced in \rref\cite{ciccarello} yields, in the continuous-time limit, our memory-kernel ME. In this model, memory effects are to be ascribed to an intraenvironmental incoherent dynamics. Next, we introduce a different CM, with no inter-ancillary collisions, for a bipartite system $S$-$M$, with $M$ undergoing collisions with the reservoir ancillas. This model mimics situations like the one encountered in the dissipative Jaynes-Cummings model. We will show that both collision models lead to the same discrete open dynamics for $S$. Moreover, in the continuous-time limit, the latter CM leads to the class of bipartite MEs that can be exactly traced over $M$ to produce our closed memory-kernel ME for $S$.

Interestingly, the idea of associating a non-Markovian dynamics with a Lindbladian dynamics on an enlarged space (obtained by adding ancillary degrees of freedom to the system) has been recently investigated \cite{budini, garrahan} (see also references therein).

This paper is organized as follows. In Section \ref{essentials} we formulate without proof our central finding, namely that a certain class of bipartite Lindbladian MEs -- whose definition is given in detail -- under partial trace over $M$ yields a memory-kernel ME for $S$. A direct proof of this result, obtained by carrying out the partial trace of the bipartite ME with the help of the Laplace transform, is given in  Section \ref{direct}. In Section \ref{review}, we illustrate a first CM  featuring inter-ancillary collisions that leads, in the continuos-time limit, to our reduced memory-kernel ME. In Section \ref{mapping}, we define a second CM describing the interaction of a bipartite system $S$-$M$ with a memoryless reservoir  (no inter-ancillary collisions) and show that the resulting discrete dynamics of $S$ coincides with the one occurring in the CM of Section \ref{review}.   Accordingly, in the continuous-time limit such reduced dynamics is governed by the memory-kernel ME discussed in Sections \ref{essentials} and \ref{direct}. General comments and final conclusions are presented in Section \ref{concl}. Some technical details and proofs are given in the Appendix.

\section{the Closed memory-kernel master equation}
\label{essentials}

Let $S$ be a quantum system of arbitrary dimension whose state we will denote as ``$\rho$". A second quantum system $M$ (the ``memory") of arbitrary dimension is coupled to $S$. Let $\hat H_{S\!M}$ be the total $S$-$M$ Hamiltonian. $M$ is additionally in contact with a bath  such  that the evolution of the joint state $\rho_{S\!M}(t)$ is governed by the Kossakowski-Linbdlad ME (we set $\hbar\ug1$ throughout)
\begin{equation}
\frac{{\rm d}\rho_{S\!M}}{{\rm d}t}=-i[\hat{H}_{S\!M},\rho_{S\!M}]+\Gamma\mathcal L_{\!M}[\rho_{S\!M}]\,, \label{ME1}
\end{equation}
where the (dimensionless) Lindblad superoperator $\mathcal L_M$ is given by
\begin{equation}
\mathcal L_M[\cdot\!\cdot\!\cdot]=\sum_{\mu\nu} \left(\hat{L}_{\mu\nu}\cdot\!\cdot\!\cdot\hat{L}_{\mu\nu}^\dag\meno\frac{1}{2}[\hat L_{\mu\nu}^\dag\hat L_{\mu\nu},\cdot\!\cdot\!\cdot]_+\right)\,, \label{lindbladian}
\end{equation}
where $[\cdot\!\cdot\!\cdot,\cdot\!\cdot\!\cdot]_+$ denotes the anticommutator (the reason for using a double index will become clear soon). In \eq (\ref{lindbladian}), the jump operators $\hat L_{\mu\nu}$ act on the Hibert space of $M$ only, as emphasized by the subscript ``$M\,$" in $\mathcal L_{\!M}$. 
Physically, \eq(\ref{ME1}) describes the situation in which $S$ is coherently coupled to $M$ while the latter is in contact with a Markovian bath. Accordingly, while the joint $S$-$M$ dynamics is Markovian, the reduced dynamics of $S$ is in general non-Markovian. A paradigmatic instance is the well-known dissipative Jaynes-Cummings model \cite{JC}, where a two-level atom (embodying $S$) is coupled to a lossy cavity mode (embodying $M$). 

While -- even in cases as (relatively) simple as the aforementioned dissipative Jaynes-Cummins model -- a partial trace over the $M$'s degrees of freedom of \eq(\ref{ME1}) does not lead to a {\it closed} ME for $\rho(t)\ug{\rm Tr}\{\rho_{S\!M}(t)\}$ \cite{nota-closed}, here, we present a class of MEs of the form (\ref{ME1}) where a closed and exact ME for $\rho(t)$ is found.

Let the initial $S$-$M$ state be a product state of the form
\begin{equation}
\rho_{S\!M}(0)=\rho_0\otimes\bar\eta_M\label{initial}\,
\end{equation}
where $\rho_0$ and $\bar\eta_M$ are arbitrary states of $S$ and $\!M$ respectively (tensor product symbols will be at times omitted in the remainder of this work).
Also, let $\eta_M$ be an arbitrary state of $M$ [in general different from $\bar\eta_M$ in \eq(\ref{initial})], which we express in a diagonal form in terms of its eigenstates $\{|\nu\rangle_M\}$ as
\begin{equation}
\eta_M=\sum_\nu p_{\nu} |\nu\rangle_M\langle \nu|\,,\label{rhoA0}
\end{equation}
where the probabilities $\{p_{\nu}\}$ are normalized ($\sum_\nu p_{\nu}\ug1$).
 
We will focus on the class of MEs (\ref{ME1}) defined by the Lindbladian superoperators $\mathcal L_{M}$ with associated jump operators
\begin{equation}
\hat L_{\mu\nu}= \sqrt{p_{\nu}}\,\, | \nu\rangle_{M}\langle \mu|\,,\label{Amunu}
\end{equation}
where $|\mu\rangle_{\!M}$ and $|\nu\rangle_{\!M}$ are generic  eigenstates of $\eta_M$  while $p_{\nu}$ is the eigenvalue corresponding to $|\nu\rangle_{\!M}$ [\cf\eq(\ref{rhoA0})]. Note that the jump operators (\ref{Amunu})  obey the completeness relation $\sum_{\mu\nu}\hat L_{\mu\nu}^\dag\hat L_{\mu\nu}\ug\openone_{\!M}$, as immediately follows from the completeness of the eigenstates $|\nu\rangle_{\!M}$  and the normalization of the $p_{\nu}$. Note that $\mathcal L_M$ is defined in terms of the state $\eta_M$ while $\hat H_{S\!M}$ [\cf\eq(\ref{ME1})] is fully arbitrary.

In the next section, we will show that for such a class of bipartite Lindbladian MEs the reduced dynamics of $S$  obeys exactly the closed memory-kernel ME
\begin{eqnarray}
\label{ME-ciccarello}
\dot{\rho}(t) &=&\Gamma\!\int_{0}^t \!\!dt'e^{-\Gamma t'} \!\mathcal E({t'})\left[\dot{\rho}(t-t') 
 \right]\!+\!e^{-\Gamma t} \dot{{\Phi}}(t) [\rho_0]\,\nonumber\\
&&+\,\Gamma e^{-\Gamma t} \left(\,\mathcal E({t})\meno \Phi({t}\,)\right)[\rho_0]\,,
\end{eqnarray}
where $\Gamma\!\ge\!0$ is a rate while $\mathcal E(t)$ and $\Phi(t)$ are CPTP quantum maps on $S$ defined by  
\begin{eqnarray}
\mathcal E(t)[\rho]&\ug&{\rm Tr}_{M}[e^{-i\hat H_{S\!M}t}\rho\otimes \eta_M \,e^{i\hat H_{S\!M}t}]\,,\label{EpstA}\\
\Phi(t)[\rho]&\ug&{\rm Tr}_{M}[e^{-i\hat H_{S\!M}t}\rho\otimes \bar\eta_M\,e^{i\hat H_{S\!M}t}]\,.\label{EpstM}
\end{eqnarray}
ME (\ref{ME-ciccarello}) is a generalisation of the memory-kernel ME first introduced in \rref\cite{ciccarello}  (see also \rref\cite{vacchini}). Specifically, the latter is retrieved in the special case $\mathcal{E}({t})\!\equiv\!\Phi({t})$, namely 
for $\eta_M\ug\bar\eta_M$ [see \eqs(\ref{EpstA}) and (\ref{EpstM})]. In Appendix A, we will provide  a direct proof that the ME (\ref{ME-ciccarello}) entails a CPTP dynamics of $S$  for any $\Gamma\!\ge\!0$ and any pair  states $\{\eta_M,\bar\eta_M\}$ of $M$.

To illustrate the nature of the quantum channel corresponding to the ME (\ref{ME1}) and in particular to the Lindbladian $\mathcal L_M$, consider the case in which $M$ is a qubit, whose Hilbert space is spanned by the othonormal basis $\{|0\rangle_M,|1\rangle_M \}$, and $\eta_M\ug |0\rangle_{\!M}\!\langle 0|$. For the sake of simplicity, let us assume here that the $M$'s initial state and the state defining the Lindbladian [see \eqs (\ref{rhoA0}) and (\ref{Amunu})] are the same, i.e., $\bar\eta_M\ug\eta_M$. According to \eq(\ref{Amunu}), the Lindbladian $\mathcal L_{\!M}$ is then defined by  the pair of jump operators
\begin{equation}
\hat L_{00}\ug|0\rangle\!_{M }\!\langle 0|\ug\frac{\hat\sigma_{Mz}\piu \openone_{M}}{2},\,\,\,\hat L_{10}\ug|0\rangle\!_{M }\!\langle 1|\ug\hat\sigma_{M-}\label{Amunu-qubit}\,,
\end{equation}
where, as usual, $\hat\sigma_{Mz}\ug |0\rangle\!_{M }\!\langle 0|\meno |1\rangle\!_{M }\!\langle 1|$ is a Pauli operator, $\hat\sigma_{M-}\ug\hat\sigma_{M+}^\dag\ug |0\rangle\!_{M }\!\langle 1|$ are ladder operators while $\openone_M$ is the identity. Hence, the Lindbladian $\mathcal L_{\!M}$ entering ME (\ref{ME1}) takes the explicit form
\begin{eqnarray}
\mathcal L_{\!M}[\rho_{S\!M}]&\ug&\!\left(\hat\sigma_{M-}\,\rho_{S\!M}\,\hat\sigma_{M+}\meno\frac{1}{2}[\hat\sigma_{M+}\hat\sigma_{M-},\,\rho_{S\!M}]_+\right)\nonumber\\
&&+\frac{1}{4}\!\left(\hat\sigma_{Mz}\,\rho_{S\!M}\,\hat\sigma_{Mz}\meno\rho_{S\!M}\right)\,.   \label{lindbladian-eq}
\end{eqnarray}
The system is thus subject to both  dissipation, with  jump operator $\hat \sigma_{M-}$, and  dephasing, with jump operator $\hat \sigma_{Mz}$, both acting on the auxiliary system $M$. Remarkably, the corresponding rates of these two decoherence processes must be in the definite 4\,:\,1 ratio. In the case that $S$ is also a qubit coupled to $M$   via an XX-type interaction Hamiltonian (this assigns the form of Hamiltonian $\hat H_{S\!M}$), the resulting ME (\ref{ME-ciccarello}), including its exact solution, has been studied in \rref\cite{phys-scrip}.

In Section \ref{direct}, we provide a direct proof that ME (\ref{ME-ciccarello}) exactly describes the reduced dynamics of $S$ entailed by \eq(\ref{ME1}) when $\mathcal L_M$ is given by \eq(\ref{Amunu}).

\section{Direct proof}\label{direct}

\subsection{The master equation in the Laplace space}
For the sake of notation compactness, let us define the superoperator $\mathcal H_{S\!M} [\cdot\!\cdot\!\cdot]\ug[\hat H_{S\!M},\cdot\!\cdot\!\cdot]$ so that ME (\ref{ME1}) can be written as
\begin{equation}
\dot{\rho}_{S\!M}\ug-i\mathcal{H}_{S\!M} [\rho_{S\!M}]\!+\!\Gamma {\mathcal{L}}_{M}[\rho_{S\!M}]\,\label{eq-rhos1}
\end{equation}
with the initial condition (\ref{initial}).
Furthermore, let us note that, thanks to \eqs (\ref{rhoA0}) and (\ref{Amunu}), the Lindbladian (\ref{lindbladian}) transforms an arbitrary joint state of $S$ and $M$ as 
\begin{equation}
\mathcal L_M[\rho_{S\!M}]={\rm Tr}_{\!M}\{\rho_{S\!M}\}\eta_M-\rho_{S\!M}\,.\label{L1}
\end{equation}

Let $\tilde{\rho}_{S\!M}(s)$ be the Laplace transform (LT) of $\rho_{S\!M}(t)$, where $s$ lies on the complex plane. Taking the LT of both sides of \eq(\ref{eq-rhos1}) and 
replacing ${\mathcal L}_M$ with \eq(\ref{L1}), we find
\begin{eqnarray}
s\tilde\rho_{S\!M}(s) \meno\rho_{S\!M}(0)&\ug&-i\mathcal H_{S\!M} [\rho_{S\!M}(s)]\nonumber\\
&&+ \Gamma\!\left[ {\rm Tr}_{M}\{\tilde\rho_{S\!M}(s)\}\,\eta_M\meno  \tilde\rho_{S\!M}(s)\right]\label{eq-phi-2}\,,
\end{eqnarray}
which can be viewed as a special case, for $\Gamma_1\ug\Gamma$, of the following, more general,  equation
\begin{equation}
(s\piu\Gamma \piu i \mathcal H_{S\!M})[\tilde{\rho}_{S\!M}(s)]\meno\rho_{S\!M}(0)\ug\Gamma_1\, {\rm Tr}_{M}\{\rho_{S\!M}(s)\}\,\eta_M\label{eq-phi-3}\,
\end{equation}
under the same initial condition (\ref{initial}).
The solution of \eq\eqref{eq-phi-3}, which depends on both $\Gamma$ and $\Gamma_1$, for a given value of $\Gamma$,
 can be expanded in powers of $\Gamma_1$ as
\begin{equation}
\tilde{\rho}_{S\!M}(s)=\sum_{k=1}^\infty\Gamma_1^{k-1} \tilde{\rho}_{S\!M}^{(k)}(s)\label{expansion}
\end{equation}
where each  $\tilde{\rho}_{S\!M}^{(k)}(s)$ parametrically depends on $\Gamma$. The solution of our \eq\eqref{eq-phi-2} can thus be obtained 
by evaluating the $\tilde{\rho}_{S\!M}^{(k)}(s)$ and then setting $\Gamma_1\ug\Gamma$ in \eq(\ref{expansion}). 

To determine $\tilde{\rho}_{S\!M}^{(k)}(s)$, we replace expansion (\ref{expansion}) in \eq(\ref{eq-phi-3}) so as to end up with the set of equations (one for each power of $\Gamma_1$)
\begin{eqnarray}\label{system}
\begin{cases}
(s\piu\Gamma \piu i \mathcal H_{S\!M})[\tilde{\rho}_{S\!M}^{(1)}(s)]&\ug \rho_{S\!M}(0)\,,\\
(s\piu\Gamma \piu i \mathcal H_{S\!M})[\tilde{\rho}_{S\!M}^{(k)}(s)]&\ug {\rm Tr}_{M}\left\{\tilde\rho_{S\!M}^{(k-1)}(s)\right\}\!\eta_M\,\,\,(k\ge2)\,.\nonumber
\end{cases}
\end{eqnarray}
The first equation (corresponding to the 0th power in $\Gamma_1$) immediately yields
\begin{equation}
\tilde{\rho}_{S\!M}^{(1)}(s)= \tilde{\mathcal U}_{S\!M}(s+\Gamma)\left[\rho_{S\!M}(0)\right]\,,\label{phi1}
\end{equation}
where 
\begin{equation}
\tilde{\mathcal U}_{S\!M}(s)\ug(s+i\mathcal H_{S\!M})^{-1}\,.\label{phi-def}
\end{equation}
Note that $\tilde{\mathcal U}_{S\!M}(s)$ is the LT of the quantum map 
\begin{equation}
\mathcal U_{S\!M}(t)\ug e^{-i \mathcal H_{S\!M} t}\,,\label{mapU}
\end{equation}
namely the unitary dynamical map on $S$-$M$ corresponding to the  ME (\ref{ME1}) [or equivalently \eq(\ref{eq-rhos1})] for $\Gamma\ug0$, i.e., in the absence of interaction with the reservoir. 
Furthermore, for $k\!\ge\!2$, we have 
\begin{equation}
\tilde{\rho}_{S\!M}^{(k)}(s)= \tilde{\mathcal U}(s+\Gamma)\left[{\rm Tr}_{M}\!\left\{\rho_{S\!M}^{(k-1)}(s)\right\}\eta_M\right]\,,\label{phi2}
\end{equation}
which is a recurrence relation allowing to determine each $\tilde{\rho}_{S\!M}^{(k)}(s)$. Replacing these in \eq(\ref{expansion}) and setting $\Gamma_1\ug\Gamma$, we thus get the solution of ME \eqref{eq-phi-2} in the Laplace space. 

\subsection{The reduced dynamics of $S$}

We now derive the reduced dynamics of $S$, i.e., we evaluate  $\rho(t)\ug {\rm Tr}_{M}\{\rho_{S\!M}(t)\}$. Let $\tilde \rho(s)$ be the LT of $\rho(t)$, from \eq\eqref{expansion}, it follows
\begin{equation}
\tilde{\rho}(s)=\sum_{k=1}^\infty\Gamma^{k-1} {\rm Tr}_{M}\left\{\tilde{\rho}_{S\!M}^{(k)}(s)\right\}\label{expansion-rhos}
\end{equation}
with $\tilde{\rho}_{S\!M}^{(k)}(s)$ given by \eqs(\ref{phi1}) and (\ref{phi2}). 

We will additionally need the LTs of maps $\mathcal E(t)$ and $\Phi(t)$ on $S$, which in the light of \eqs(\ref{EpstA}) and (\ref{EpstM}) can be expressed in terms of the joint map (\ref{phi-def}) as
\begin{eqnarray}
\mathcal {\tilde E}(s)[\rho]&\ug&{\rm Tr}_{M}\left\{\tilde{\mathcal U}_{S\!M}(s)[\rho\otimes \eta_M] \,\right\}\,,\label{EpstA2}\\
\tilde\Phi(s)[\rho]&\ug&{\rm Tr}_{M}\left\{\tilde{\mathcal U}_{S\!M}(s)[\rho\otimes \bar\eta_M] \right\}\,.\label{EpstM2}
\end{eqnarray}

For $k\ug1$, recalling the initial condition \eq(\ref{initial}), we thus find from \eq\eqref{phi1} 
$$ {\rm Tr}_{M}\left\{\tilde{\rho}_{S\!M}^{(1)}(s)\right\}\ug \tilde{\Phi}(s+\Gamma)[\rho_0]\,.$$

For $k\ug2$, from \eqs(\ref{phi2}) and (\ref{EpstA2}) it immediately follows
$$ {\rm Tr}_{M}\left\{\tilde{\rho}_{S\!M}^{(2)}(s)\right\}\ug \tilde{\mathcal E}(s+\Gamma)\comp\tilde{\Phi}(s+\Gamma)\,[\rho_0]\,.$$
Likewise, for $k\ug3$
$$ {\rm Tr}_{M}\left\{\tilde{\rho}_{S\!M}^{(3)}(s)\right\}\ug\left[\mathcal E(s+\Gamma)\right]^2\comp\tilde{\Phi}(s+\Gamma)\,[\rho_0]\,$$
with $[\tilde{\mathcal E}(s+\Gamma)]^2\ug\tilde{\mathcal E}(s+\Gamma)\!\comp\!\tilde{\mathcal E}(s+\Gamma)$.
By induction,  for arbitrary $k\!\ge\!1$
$$ {\rm Tr}_{M}\left\{\tilde{\rho}_{S\!M}^{(k)}(s)\right\}\ug\tilde{\mathcal E}^{k-1}(s+\Gamma)\comp\tilde{\Phi}(s+\Gamma) [\rho_0]\,.$$
with $\tilde{\mathcal E}^{0}\ug \mathbb{I}_S$.
Substituting into \eq(\ref{expansion-rhos}), the solution for $\rho(s)$ thus reads
\begin{eqnarray}
\tilde{\rho}(s)&\ug&\sum_{k=1}^\infty\Gamma^{k-1}\! \tilde{\mathcal E}^{k-1}(s+\Gamma)\comp\tilde{\Phi}(s+\Gamma)[\rho_0]\nonumber\\
&&\ug \left[\mathbb{I}_S-\Gamma \,\tilde{\mathcal E}(s+\Gamma)\right]^{-1}\!\!\comp\tilde{\Phi}(s+\Gamma)\,\,[\rho_0]\,,\label{solution}
\end{eqnarray}
It is straightforward to show (see Appendix A) that in the Laplace representation this precisely coincides with the solution of ME (\ref{ME-ciccarello}) under the initial condition (\ref{initial}). This shows that MEs (\ref{ME1}) and (\ref{ME-ciccarello}) correspond to the same open dynamics of $S$, which completes our proof.

In the remainder of this paper, we will show how the steps leading from MEs (\ref{ME1}) to (\ref{ME-ciccarello}) can be described and understood in terms of CMs in the continuous time limit.

\section{A collision Model with internal memory}\label{review}

In collision models the reservoir $R$ consists of a large collection of identical ancillas -- here labeled with a positive integer $n\ug1,2,...$ -- all initially in the same state $\eta$. The interaction between a system $S$ and $R$  is described in terms of a sequence of collisions, each lasting a short time $\tau$,  between $S$ and the $n$th ancella. Each collision is described by a unitary operator $\hat U_{Sn}\ug e^{-i \hat H_{Sn}\tau}$ where $\hat H_{Sn}$ is the interaction Hamiltonian. The Hilbert space dimension of both $S$ and a generic ancilla can be arbitrary.  The initial - product - state of $S$-$R$ reads 
\begin{equation}
\sigma_0\ug\rho_0\!\otimes\!\eta_1\!\otimes\!\eta_2\!\otimes...\,\label{sigma0}
\end{equation} 
where $\rho_0$ is the initial state of $S$. In the absence of any other type of dynamical processes, at the $n$th step the joint $S$-$R$ state is
\begin{equation}
\sigma_n\ug(\hat U_{Sn}\!\cdot\!\cdot\!\cdot\!\hat U_{S2}\hat U_{S1})\rho_0\!\otimes\!\eta_1\!\otimes\!\eta_2\!\otimes...\eta_n(\hat U_{Sn}\!\cdot\!\cdot\!\cdot\!\hat U_{S2}\hat U_{S1})^\dag\,.
\end{equation}
Provided the system collides only once with the same ancilla this leads to a fully Markovian open dynamics for $S$. Indeed, if $\rho_n\ug {\rm Tr}_R\{\sigma_n\}\!\equiv\!{\rm Tr}_n\{\sigma_n\}$, at the $n$th-step: $\rho_n\ug \mathcal E_\tau [\rho_{n-1}]\ug \mathcal E_\tau^n[\rho_0]$, where map $\mathcal E_\tau\ug \mathcal E(\tau)$ coincides with \eq(\ref{EpstA}) for $t\ug\tau$ and $\eta \!\equiv\!\eta_M$. The above would give rise to a standard {\it memoryless} CM \cite{CMs,scarani} (note that the discrete dynamical map $\mathcal E_\tau^n$ fulfils the semigroup property), which in the continuous-time limit gives rise to a Lindblad-type ME for $S$ \cite{buzekME}.

\begin{figure}[h!]
\begin{center}
\includegraphics[width=8.7cm]{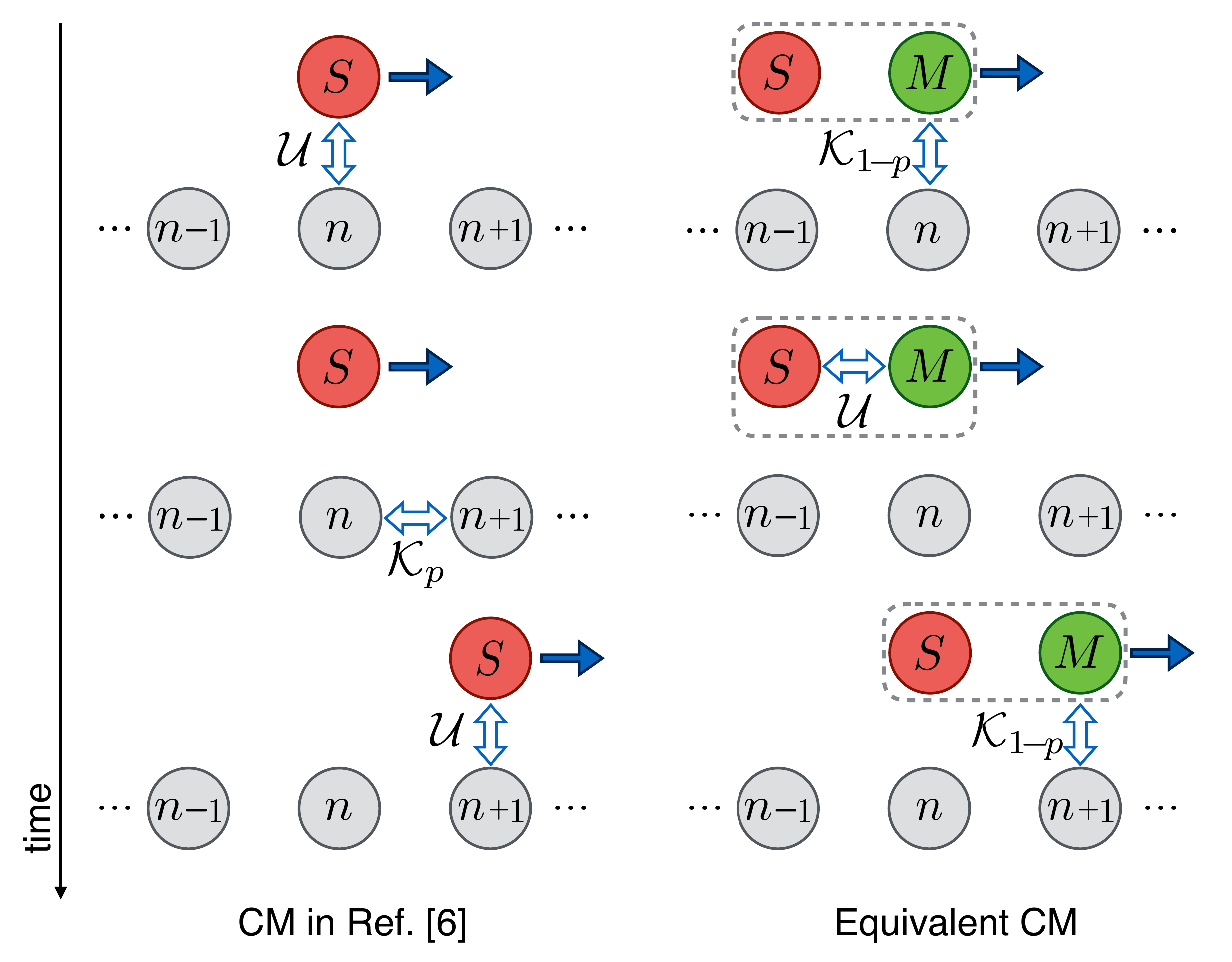}
\caption{(a) CM with internal memory: inter-ancillary collisions are interspersed with system-ancilla collisions. (b) Equivalent Markovian collision model for $S$-$M$: only the auxiliary system $M$ undergoes successive interactions with the bath ancillas, which are interspersed with unitary $S$-$M$ collisions. In the latter model no inter-ancillary collisions occur. \label{fig1}}
\label{NM}
\end{center}
\end{figure}

A way to endow the open dynamics of $S$ with memory is to introduce {\it inter-ancillary} pairwise collisions occurring  between consecutive system-ancilla (SA) interactions.  In other words, the collision between $S$ and the $n$th ancilla is followed by a collision between the $n$th and $(n\piu1)$th ancillas, which is in turn followed by a new collision involving $S$ and the $(n\piu1)$th ancilla, then by a new   collision between ancillas $n\piu1$ and $n\piu2$, and so on. A pictorial sketch of such dynamics is given in \fig\ref{fig1}(a).
In line with \rref \cite{ciccarello} the  inter-ancillary collision between ancillas $n$ and $n\piu1$ is modelled as a probabilistic swap operation, which is a non-unitary process with an associated quantum map that transforms the joint $S$-$R$ state $\sigma$ according to
\begin{equation}
\mathcal K^{(n,n+1)}_{p}[\sigma]\ug   q  \sigma\piu p \; \hat{S}_{n,n+1}\sigma\hat{S}_{n,n+1}\,\,\,\,\,\,\,\,\,\,\,(q\ug 1\meno p)\,\,,\label{mapA}
\end{equation}
where $\hat{S}_{i,j}$ is the swap unitary operator exchanging the states of ancillas $i$ and $j$. In other words, the states of the two involved ancillas are swapped with probability $p$ or left unchanged with probability $q$. We define as ``step" the product of an inter-ancillary and of the following system-ancilla collisions.  Besides introducing inter-ancillary collisions, we also slightly generalise the initial state (\ref{sigma0}) as
\begin{equation}
\sigma_0\ug\rho_0\!\otimes\!\bar\eta_1\!\otimes\!\eta_2\!\otimes\!\eta_3\!\otimes...\,,\label{sigma0-gen}
\end{equation} 
where the initial state of ancilla 1, $\bar\eta_1$, is in general different from the common state of ancillas $\eta_n = \eta$ with $n\!\ge\!2$. The reason for considering this more general state will  become clear later.

After the first $S$-1 collision, the initial joint state $\sigma_0$ [\cf\eq(\ref{sigma0-gen})] is transformed into $\sigma_1\ug\hat{U}_{S\!1}^\dag\sigma_0\hat U_{S\!1}^\dag$. Next, the inter-ancillary 1-2 collision occurs followed by the  collision $S$-2, which yields, at the end of the 2nd step, $\sigma_2\ug q (\hat U_{S2}\sigma_1\hat U_{S2}^\dag)\piu p (\hat U_{S2}\hat S_{12}\sigma_1\hat S_{12}\hat U_{S2}^\dag)$. This can equivalently be expressed in the form
\begin{eqnarray}
\sigma_2 \ug q (\hat U_{S2}\sigma_1\hat U_{S2}^\dag)\piu p\,\hat{U}_{S2}^2 (\rho_0\!\otimes\!\eta_1\!\otimes\!\bar\eta_2\!\otimes\!\eta_3\!\otimes...)\hat{U}_{S2}^{\dag2}\label{sigma2-bis}\,,\,\,\,\,\,
\end{eqnarray}
where in the second term we replaced $\sigma_1{=}\hat U_{S\!1}\sigma_0\hat U_{S\!1}^\dag$ and used the identity $\hat S_{12}\hat{U}_{S\!1}{=}\hat{U}_{S2}\hat S_{12}$. Accordingly, the state between brackets in the second term of \eq (\ref{sigma2-bis}) is the initial state (\ref{sigma0-gen}) where ancillas 1 and 2 have been swapped (note that state $\bar\eta$ is now the state of ancilla 2).
By iteration (see Appendix B), the state at the $n$th step will read
\begin{eqnarray}\label{sigman}
\sigma_n&\ug&  q \sum_{j=1}^{n\meno1}p^{j-1}{\hat U}_{Sn}^j \sigma_{n\meno j}{\hat U}_{Sn}^{j\dag} \nonumber\\
&\piu& \;  p^{n-1} {\hat U}^n_{Sn} (\rho_0\!\otimes\!\eta_1\!\otimes...\otimes\eta_{n-1}\!\otimes\!\bar\eta_n\!\otimes\eta_{n+1}\!\otimes...){\hat U}_{Sn}^{n\dag}.\,\,\,\,\,\,\,\,\,\,
\end{eqnarray}
In the state between brackets in the second term, the $n$th ancilla is in state $\bar\eta$, while all the remaining ones are in $\eta$. Note that in \eq(\ref{sigman}) only the unitary operator ${\hat U}_{Sn}$ associated with the $n$th SA collision appears. This remarkable property allows to write the corresponding equation for the {\it reduced} $S$ density operator $\rho_n\ug{\rm Tr}_R\sigma_n$ as
\begin{equation}
\rho_n\ug q \sum_{j=1}^{n\meno1}p^{j-1}\mathcal{E}_j[\rho_{n\meno j}]\piu  p^{n-1} \Phi_n[\rho_0]\,.\label{rhon-gen}
\end{equation}
where [\cf \eqs(\ref{EpstA}) and (\ref{EpstM})] $\mathcal{E}_j\!\equiv\! \mathcal E(j\tau)$ and $\Phi_j\ug \Phi(j\tau)$. 
Correspondingly, the variation of $\rho_n$ between the $(n\meno1)th$ and $n$th steps, i.e., $\Delta \rho_n\ug (\rho_n\meno\rho_{n-1})$, is given by
\begin{eqnarray}
\!\Delta \rho_n&\ug& q \!\sum_{j=1}^{n\meno2}p^{j-1}\mathcal{E}_j[\Delta \rho_{n\meno j}]\piu q\, p^{n\meno1} \mathcal{E}_{n\meno1}[\rho_1]
\piu \Delta\! \left(p^{n\meno1} \Phi_n\right)\!\![\rho_0].\!\!\!\!\nonumber\\\label{drhon-gen}
\end{eqnarray}
In the continuous-time limit, \eq(\ref{drhon-gen}) can be cast in the form of an exact ME. In this limit $n\tau\!\rightarrow\! t\,$ and $j\tau\!\rightarrow\! t'$ so that $p^{j}\ug ({p^{\frac{1}{\tau}}})^{j\tau}\ug e^{-\Gamma t'}$,  where the memory rate $\Gamma$ is defined in terms of $p$ [\cf\eq(\ref{mapA})] and $\tau$ as 
\begin{equation}
p= \exp[-\Gamma \tau]\label{p-Gamma}\,.
\end{equation}
Furthermore, in the same limit, $\tau$ must be far shorter than any characteristic time, in particular $\Gamma^{-1}$. Hence, $\Gamma\tau\!\ll\!1$ and thus $q\ug 1\meno p\ug1\meno e^{-\Gamma \tau}\!\simeq\!\Gamma \tau$. When this is used in \eq(\ref{drhon-gen}), we end up with the memory-kernel ME (\ref{ME-ciccarello}) as we show in detail in Appendix C.

This collision model generalizes the one in \rref\cite{ciccarello}, the latter being retrieved in the special case $\bar \eta\equiv\eta$. Such an extension is indeed necessary to ensure that the closed memory-kernel ME (\ref{ME-ciccarello}) corresponds to a bipartite ME for $S$-$M$ where the memory $M$ is initially in an {\it arbitrary} state $\bar \eta$ (see Section \ref{essentials}). In the CM discussed above, this requirement simply translates into allowing the 1st and the remaining reservoir ancillas to be initially in different states.

\section{A memoryless collision model}\label{mapping}

We now show that, as anticipated in the introduction, our memory-kernel ME can be derived by a second memoryless CM describing a subsystem $S$ coupled to a fully Markovian environment via an auxiliary system $M$. In this CM the reservoir $R$ consists again  of a large collection of identical ancillas $n\ug1,2,...$, which however are now non-interacting (no inter-ancillary collisions occur). The ``system" relaxing into the reservoir $R$ is now {\it bipartite}, with its subsystems  $S$ and $M$ mutually interacting according to a Hamiltonian $\hat H_{S\!M}$. By hypothesis, the Hibert-space dimensions of $M$ and each of the ancillas are assumed to be equal. The initial $S$-$M$-$R$ joint state is  assumed to be
\begin{equation}
\sigma_0=\left[\rho_0\otimes\bar\eta_M\right]\otimes\!\eta_1\!\otimes\!\eta_2\otimes\cdot\!\cdot\!\cdot\,\,\,.\label{sigma0-bi}
\end{equation}
As the reservoir ancillas do not interact with each other (in contrast to the CM of the previous section) here only system-ancilla collisions take place.
These collisions, which we assume to involve only $M$ (i.e., $S$ is not in direct contact with $R$) are described by the {\it non-unitary} probabilistic swap $\mathcal K^{(M,n)}_{1-p}$. The definition of such a map, acting on $M$ and the $n$th ancilla, is the same as in \eq(\ref{mapA}) {\it but the replacement} $p\!\rightarrow\!1\meno p$ (as highlighted by the subscript). In other words, at each collision, either the current state of $M$ is swapped with the ancillary state $\eta$ with probability $1\meno p$ or left unchanged with probability $p$. In addition to such collisions a unitary dynamics internal to the bipartite $S$-$M$ system, generated by the Hamiltonian $\hat H_{S\!M}$, takes place. Such dynamics has the form of $S$-$M$ unitary collisions that are interspersed with collisions between $M$ and the reservoir ancillas. A sketch of the CM dynamics is shown in \fig1(b). Note that the joint $S$-$M$ system undergoes a fully Markovian dynamics (while in general this is not the case for $S$).

It is convenient to define a map $\mathcal S$ on $S$-$M$ as
\begin{eqnarray}
\mathcal S(\rho_{S\!M})&=&\Tr_{n}\left\{\hat S_{Mn}(\rho_{S\!M}\,\eta_n)\hat S_{Mn}^\dagger\right\}={\rm Tr}_M\{\rho_{S\!M}\}\eta_{M},\,\,\,\,\,\,\,\,\,\label{map-S}
\end{eqnarray}
which describes how an arbitrary $S$-$M$ state $\rho_{S\!M}$ is changed after that a (unitary) swap operation is applied on $M$ and a generic ancilla initially in state $\eta$ (we recall that this has the same dimension as $M$). A proof of the last identity in \eq(\ref{map-S}) is given in Appendix D.
Using \eq(\ref{map-S}), from \eq(\ref{mapA}) we get
\begin{equation}
{\rm Tr}_n\left\{\mathcal K^{(M,n)}_{1-p}[\rho_{S\!M}\eta_n]\right\}\ug   p \, \rho_{S\!M}+ q\, \mathcal S[\rho_{S\!M}]\,.\label{mapA2}
\end{equation}

Initially (0th step), $S$ and $M$ are in the state $\rho_{S\!M}^{(0)}\ug \rho_0 \bar\eta_M$, each of the ancillas being in state $\eta$ [see \eq(\ref{sigma0-bi})].
They then collide with each other, hence their state after the 1st step reads $\rho^{(1)}_{S\!M}\ug\U[\rho^{(0)}_{S\!M}]$. We have set for brevity $\mathcal U_\tau\ug \mathcal U_{S\!M}(\tau)$ [\cf\eq(\ref{mapU})].
Next, an $M$-1 collision described by map $\mathcal K_{1-p}^{(M,1)}$ takes place followed by a new $S$-$M$ unitary collision. Hence, at the 2nd step
\begin{eqnarray}
\rho^{(2)}_{S\!M}&\ug&p~\U\left[\rho^{(1)}_{S\!M}\right]\!+\! q ~\U\!\comp\mathcal S \left[\rho^{(1)}_{S\!M}\right]\nonumber\\		      
&\ug& p~\U^2\left[\rho^{(0)}_{S\!M}\right]\!+\! q ~\U\mathcal S \left[\rho^{(1)}_{S\!M}\right]\label{rhoS1-2}\,
\end{eqnarray}
(the map composition symbol ``$\comp$" will be always omitted henceforth).

At the 3rd step,
\begin{eqnarray}
\rho^{(3)}_{S\!M}&=&p~\U\left[\rho^{(2)}_{S\!M}\right]\!+\! q ~\U\mathcal S \left[\rho^{(2)}_{S\!M}\right]  \nonumber\\
		      &=&p~\U\!\left[p~\U^2\left[\rho^{(0)}_{S\!M}\right]\!+\! q ~\U\mathcal S \left[\rho^{(1)}_{S\!M}\right]\right]\!+\! q ~\U\mathcal S \left[\rho^{(2)}_{S\!M}\right]\nonumber\\
     		      &=&p^2~\U^3\left[\rho^{(0)}_{S\!M}\right]\!+\! q p~\U^2\mathcal S \left[\rho^{(1)}_{S\!M}\right]\!+\! q ~\U\mathcal S \left[\rho^{(2)}_{S\!M}\right]\,,\nonumber
\end{eqnarray}
where to obtain the second identity we have used \eq(\ref{rhoS1-2}).

By induction (see Appendix \ref{app-ind}), the $n$th-step state can be arranged as
\begin{equation}
\rho^{(n)}_{S\!M}\ug q \sum_{j=1}^{n-1}p^{j-1}~\U^j\,\mathcal S \left[\rho^{(n-j)}_{S\!M}\right]\!+\!p^{n-1}\U^n\!\left[\rho^{(0)}_{S\!M}\right]\label{rhoS1-n}\,.
\end{equation}
Due to \eq(\ref{map-S}), $\mathcal S [\rho^{(n-j)}_{S\!M}]\ug\rho_{n-j}\,\eta_{M}$. Replacing this and $\rho_{S\!M}(0)\ug\rho_{0}\,\bar\eta_{M}$ in \eq(\ref{rhoS1-n}), upon trace over $M$ we thus find
\begin{equation}
\rho^{(n)}_{S\!M}\ug q \sum_{j=1}^{n-1}p^{j-1}~\U^j \left[\rho_{n-j}\,\eta_{M}\right]+p^{n-1}\U^n\left[\rho_{0}\,\bar\eta_M\right]\,.
\end{equation}
Upon trace over $M$, recalling the definition of maps $\mathcal E(t)$ and $\Phi(t)$ [\cf\eqs(\ref{EpstA}) and (\ref{EpstM})], we thus end up with \eq(\ref{rhon-gen}).
This shows that, as far as the reduced dynamics of $S$ is concerned, the present CM, in which the joint $S$-$M$ system undergoes incoherent binary collisions between $M$ and each of the non-interacting ancillas of an infinitely large reservoir interspersed with ``internal" $S$-$M$ coherent unitary collisions, is equivalent to the  
discrete CM described in Section \ref{mapping}.

\subsection*{Microscopic derivation of Master Equation (\ref{ME1})}
We now show that, in the continuous-time limit, the bipartite CM discussed above gives rise to ME (\ref{ME1}). To this end, we use an approach similar to the one adopted for composite CMs \cite{composite} (these differ from the present CM in that, unlike here, the system-ancilla collisions are unitary). 
At step $n$, the $S$-$M$ joint state is
\begin{eqnarray}
\rho_{S\!M}^{(n)}&=& p {\rm Tr}_{n} \!\left\{e^{-i\hat H_{S\!M}\tau}\rho_{S\!M}^{(n-1)}\eta_n e^{i\hat H_{S\!M}\tau}\right\}\nonumber\\
&&+ q {\rm Tr}_{n} \!\left\{e^{-i\hat H_{S\!M}\tau}\hat S_{M\!n}\,\rho_{S\!M}^{(n-1)} \eta_n  \hat S_{M\!n}e^{i\hat H_{S\!M}\tau}\right\},\,\,\,\,\,\,\,\label{rhoSMME}
\end{eqnarray}
where $p$ is given by \eq(\ref{p-Gamma}) and we have replaced the explicit forms of maps (\ref{mapU}) and (\ref{map-S}). 

In the continuous-time limit [see also the discussion following \eq(\ref{p-Gamma})], $\tau\!\simeq\!0$ in a way that $p\!\simeq\!1\meno\Gamma\tau$ and $q\!\simeq\!\Gamma\tau$. Hence, up to first order in $\tau,$ $e^{-i\hat H_{S\!M}\tau}\rho_{S\!M}^{(n-1)} \eta_n e^{i\hat H_{S\!M}\tau}\!\simeq\!\rho_{S\!M}^{(n-1)}\eta_n \meno i\tau[\hat H_{S\!M},\rho_{S\!M}^{(n-1)} \eta_n]$. An analogous result is obtained [see second line of \eq(\ref{rhoSMME})] under the replacement $\rho_{S\!M}^{(n-1)} \eta_n\!\rightarrow\!\hat S_{M\!n}\rho_{S\!M}^{(n-1)} \eta_n\hat S_{M\!n}$ Thereby, the $n$th-step state change 
$\Delta \rho_{S\!M}^{(n)}\ug\rho_{S\!M}^{(n)}\meno\rho_{S\!M}^{(n-1)}$, up to first order in $\tau$, takes the form
\begin{eqnarray}
\Delta \rho_{S\!M}^{(n)}&\ug&\meno i \tau\left[\hat H_{S\!M},\rho_{S\!M}^{(n)}\right] \meno\Gamma\tau \rho_{S\!M}^{(n-1)}\nonumber\\
&&+\Gamma \tau\, {\rm Tr}_{n} \!\left\{\hat S_{M\!n}\,\rho_{S\!M}^{(n-1)}\eta_n \hat S_{M\!n}\right\}\,.\label{drhon-swap}
\end{eqnarray}
In the continuos-time limit, $\Delta \rho_{S\!M}^{(n)}/\tau\!\rightarrow\!\dot \rho_{S\!M}$. Using this and the last identity in \eq(\ref{map-S}) we thus end up with ME (\ref{ME1}) [we recall that the Lindbladian $\mathcal L_M$ 
defined by \eq(\ref{Amunu}) can be equivalently expressed in the form (\ref{L1})].

We conclude by pointing out that upon decomposition of the ancillary state $\eta_n$ in its eigenstates [see \eq(\ref{rhoA0}) for $M\!\rightarrow\!n$] the partial trace in \eq(\ref{drhon-swap}) can be expressed as
\begin{eqnarray}
{\rm Tr}_{n}\! \left[\hat S_{M\!  n}\rho_{S\!M}^{(n-1)}\eta_n \hat S_{M\!  n}\right]\ug{\sum_{\mu}} {_n}\!\bra{\mu}\hat S_{M\!  n}\,\rho_{S\!M}^{(n-1)}\eta_n \hat S_{M\!  n}\ket{\mu}_n\ug\nonumber\\
\sum_{\mu\nu} p_\nu\,{_n}\!\bra{\mu} \hat S_{M\!  n}\ket{\nu}_n \rho_{S\!M}^{(n-1)}\bra{\nu}_n\hat S_{M\!  n}\ket{\mu}_n\ug
\sum_{\mu\nu} \hat L_{\mu\nu}\rho_{S\!M}^{(n-1)}\hat L_{\mu\nu}^\dagger\,,\nonumber
\end{eqnarray}
where $\{\hat L_{\mu\nu}\}$ indeed coincide with the jump operators (\ref{Amunu}). This illustrates that the form of the Lindbladian $\mathcal L_M$ introduced in Section \ref{essentials} can 
be interpreted as stemming from swap-like interactions between $M$ and the reservoir $R$.

Note that the above can be regarded as an  ``indirect" demonstration of the fact that ME (\ref{ME-ciccarello}) arises from ME (\ref{ME1}) upon trace over $M$.

\section{Conclusions}\label{concl}

In this work, we addressed the problem of an open quantum system $S$ coherently coupled to a memory $M$, which in turn is in contact with a Markovian bath, the bipartite $S$-$M$ system being governed by a Lindblad-type ME. In contrast to the typical case where tracing out the degrees of freedom of $M$ does not yield a closed ME, we have found a class of MEs of the above form where this partial trace does give rise to a closed, exact ME for $S$. This can be viewed as a generalisation of a ME originally derived via a CM. This lead us to interpret the link between the memory-kernel ME for $S$ and the $S$-$M$ Lindbladian ME in terms of suitably defined CMs, hence providing in fact a comprehensive microscopic framework underlying our central result. In particular, we have shown that our  ME can be derived as the continuous-time limit of two distinct collision models each describing a different physical scenario.

We note (see also footnote \cite{nota-closed}) that in the case of the dissipative Jaynes-Cummins model, which is also governed by a bipartite Lindbladian ME where an atom (mode) embodies $S$ ($M$), it is known that a reduced ME for the atom can be obtained \cite{petruccione} but this is in fact formulated in terms of its solution (as if it were {\it a priori} known), at variance with our case. One could object that even the memory-kernel ME (\ref{ME-ciccarello}) is expressed in terms of maps $\mathcal E(t)$ and $\Phi(t)$, which are assumed to be known. Note, though, that these can be regarded as the solutions of the problem (for two different initial conditions) where $S$ and $M$ undergo a {joint} {\it unitary} evolution, fully dependent on their total Hamiltonian $\hat H_{S\!M}$, and one aims at working out the corresponding dynamical map of $S$. Such a problem is often amenable to analytical solution.

We also comment on the relationship with the ME in \rref\cite{ciccarello} and, accordingly, the associated CM. In the case $\mathcal E(t)\!\equiv\!\Phi(t)$, $\eta_M$ would coincide with $\bar\eta_M$. This would bring about that, for a given Lindbladian (\ref{lindbladian}) specified by the jump operators (\ref{Amunu}), there would be only a single possible initial state $\bar \eta_M$ of $M$ such that the partial trace of ME (\ref{ME1}) leads to ME (\ref{ME-ciccarello}). Alternatively, given an arbitrary $M$ initial state $\bar\eta_m$, only the Lindbladian specified by the jump operators (\ref{Amunu}) with $\eta_M\!\equiv\!\bar\eta_M$ would entail \eq(\ref{ME-ciccarello}) for $S$. Instead, the presence of the {two} maps  $\mathcal E(t)$ and $\Phi(t)$ in  ME (\ref{ME-ciccarello}) ensures the existence of a class of bipartite MEs such that, for any element of this, the partial trace over $M$ leads to ME (\ref{ME-ciccarello}) for {\it any} initial state of $M$ entering \eq(\ref{initial}). Furthermore, it is easy to show \cite{nota-kossa} that, as mentioned in the Introduction, the sufficient conditions found in \rref \cite{kossa} for a well-behaved Nakajima-Zwanzig ME are satisfied in our case \eq(\ref{ME-ciccarello}).

The special property of the class of MEs specified by Lindbladian (\ref{Amunu}), which enables to work out a closed ME for $S$, is that the $M$-reservoir coupling is based on swap-like interactions (the properties of the swap operator somehow enters all of our proofs). While this is a peculiar model, yet not academic [see \eq(\ref{lindbladian-eq})], we envisage that it can be exploited  as an advantageous theoretical testbed for investigating quantum non-Markovianity concepts.

\section*{Acknowledgements}
We acknowledge support from the EU Projects TherMiQ (Grant Agreement 618074) and QuPRoCs (Grant Agreement 641277).

\appendix

\section{Laplace transform of master equation (\ref{ME-ciccarello})}\label{app-A}

Here, we derive the solution of ME (\ref{ME-ciccarello}) in the Laplace space. The proof is based on calculations that are similar to those in \rref\cite{ciccarello,ciccarelloSM}. Let $\Lambda(t)$ be the dynamical map describing 
the non-unitary dynamics of $S$ corresponding to ME (\ref{ME-ciccarello}) according to $\rho(t)\ug\Lambda(t)[\rho_0]$ with $\Lambda(0)\ug\mathbb{I}_S$. 
Map $\Lambda(t)$ obeys \eq(\ref{ME-ciccarello}) under the formal replacement $\rho\!\rightarrow\!\Lambda$. 
Indeed, replacing $\rho(t) = \Lambda(t) [\rho_0]$ in \eqref{ME-ciccarello} and using that $\rho_0$ is arbitrary, we find that $\Lambda(t)$ is governed by the equation
\begin{eqnarray}\label{ME}
\dot{\Lambda}(t)&=&\Gamma\!\int_{0}^t \!\!dt'e^{-\Gamma t'} \;\mathcal{E}({t'})\left[ \dot{\Lambda}(t-t') \right]\nonumber\\
&&+\Gamma e^{-\Gamma t}({\cal E}(t)-\Phi(t))+e^{-\Gamma t}\; \dot{\Phi}(t)\;.
\end{eqnarray}
Upon LT, the equation becomes
\begin{eqnarray} \label{ltmes}
 \tilde{{\Lambda}}(s) 
\ug 
\Gamma\, \tilde{\mathcal E}(s+\Gamma)   \tilde{{\Lambda}}(s)+\; \tilde{\Phi}(s+\Gamma)
\end{eqnarray} 
where for $s$ complex the LT is defined as 
\begin{equation}
\tilde{F}(s) \ug { L}\,[F(t)](s)\ug \int_0^\infty\!\! d t \; e^{-s t} F(t)\,\label{LT}.
\end{equation} 
By rearranging terms in \eq(\ref{ltmes}) as
\begin{eqnarray} 
 [ {\mathcal I} -  \Gamma \; \tilde{\mathcal E}(s+\Gamma) ] \  \tilde{{\Lambda}}(s)=  
 \tilde{\Phi}(s+\Gamma) 
\end{eqnarray} 
and introducing the inverse of map ${\mathcal I} \meno  \Gamma \; \tilde{\mathcal E}(s+\Gamma)$ we end up with [\cf\eq(\ref{solution})]
\begin{eqnarray} \label{ffd}
 \tilde{{\Lambda}}(s)=    \left[{\mathcal I} - \Gamma\;\tilde{\mathcal E}(s\!+\!\Gamma)\right]^{-1}\!\!\!\comp \tilde{\Phi}(s\!+\!\Gamma)\;.\,\,
\end{eqnarray} 
Expanding \eq(\ref{ffd}) in powers of $\Gamma$ gives
\begin{equation}
\tilde{{\Lambda}}(s)\ug\sum_{k=1}^\infty\Gamma^{k\!-\!1}\tilde{\mathcal{E}}^{k-1}(s\!+\!\Gamma)\!\circ\tilde{\Phi}(s\!+\!\Gamma)
\end{equation}
whose inverse LT is
\begin{eqnarray} \label{ffd3}
\Lambda(t) 
\,\ug  \sum_{k=1}^\infty\Gamma^{k\!-\!1} \; { L}^{-1}\!\left[\tilde{\mathcal{E}}^{k-1}(s\!+\!\Gamma)\!\circ\tilde{\Phi}(s\!+\!\Gamma)\right](t)\;.
\end{eqnarray} 
Basic properties of LT allow to immediately calculate the inverse LT within brackets as 
\begin{eqnarray} 
&&{L}^{-1}\!\left[\tilde{\mathcal{E}}^{j}(s\!+\!\Gamma)\!\circ\tilde{\Phi}(s\!+\!\Gamma)\right]\ug \nonumber\\
&&e^{-\Gamma t}\!\!\!
\int_0^t \!\!\!dt_1
\!\cdot\!\cdot\!\cdot\!\!\!
\int_0^{t_{j\meno1}} \!\!\!\!\!\!\!d t_{j} \;
\mathcal{E}(t_1)\circ\cdot\!\cdot\!\cdot\!\circ\mathcal{E}(t_{j})\circ\Phi(t_{j-1}\meno t_{j}).\,\,\, \,
\label{ltn}
\end{eqnarray} 
The {integrand} in \eq(\ref{ltn}) is evidently a composition of CPTP quantum maps, hence it is CPTP itself [we recall that both $\mathcal E(t)$ and $\Phi(t)$ are CPTP maps, see \eqs(\ref{EpstA}) and (\ref{EpstM})]. Therefore, we see that the dynamical map \eq(\ref{ffd3}) is in fact a combination of CPTP maps with {\it positive} weights. This proves that the dynamical map $\Lambda(t)$ corresponding to ME (\ref{ME-ciccarello}) is completely positive and trace preserving.

 \section{Induction proof of \eq (\ref{sigman})}\label{iteration}

At the 3rd step, the overall state is given by $\sigma_3\ug q (\hat U_{S3}\sigma_2\hat U_{S3}^\dag)\piu p (\hat U_{S3}\hat S_{23}\sigma_2\hat S_{23}\hat U_{S3}^\dag)$, which with the help of \eq(\ref{sigma2-bis}) can be arranged as
\begin{eqnarray}
\sigma_3&\ug& q \sum_{j=1}^{2}p^{j{-}1}{\hat U}_{S3}^j \sigma_{3\meno j}{\hat U}_{S3}^{j\dag} \nonumber\\
&&\piu \;  p^2\,\hat{U}_{S3}^3 (\rho_0\!\otimes\!\eta_1\!\otimes\!\eta_2\!\otimes\!\bar\eta_3\!\otimes...)\hat{U}_{S3}^{\dag3}\nonumber
\label{sigma3-bis}\,,
\end{eqnarray}
where we used $\hat S_{23}\hat{U}_{S2}{=}\hat{U}_{S3}\hat S_{23}$ and $\hat S_{23}\sigma_1\hat S_{23}{=}\sigma_1$ (the latter identity follows from the fact that, at the end of the 1st step, ancillas 2 and 3 are still in the initial state $\eta$). \eq(\ref{sigman}) thus holds for $n\le3$. To prove that it is valid for arbitrary $n$, let us consider the state after the $(n{+}1)$th step, which reads
\begin{eqnarray}
\sigma_{n{+}1}&\ug& q \left(\hat U_{S,n+1}\sigma_n\hat U_{S,n+1}^\dag\right)\nonumber\\&&+ p \left(\hat U_{S,n+1}\hat S_{n,n{+}1}\sigma_n\hat S_{n,n{+}1}\hat U_{S,n+1}^\dag\right)\nonumber
\end{eqnarray}
Substituting \eq\eqref{sigman} in the second term yields
\begin{eqnarray}
\!\!\sigma_{n{+}1}&\ug&  q \sum_{j=0}^{n\meno1}p^{j}{\hat U}_{Sn}^{j+1}\sigma_{n\meno j}{\hat U}_{Sn}^{{j+1}\dag} \nonumber\\
&&\piu \;  p^{n} {\hat U}^{n{+}1}_{S,n+1} (\rho_0\!\otimes\!\eta_1\!\otimes\!...\!\otimes\eta_{n}\!\otimes\!\bar\eta_{n+1}\!\otimes\eta_{n+2}\!\otimes\!...\!){\hat U}_{S,n+1}^{n{+}1\dag} \,.\nonumber
\end{eqnarray}
Rearranging index $j$ in the above expression we end up with \eq\eqref{sigman} for $n\rightarrow n{+}1$. This proves that \eq\eqref{sigman} holds.

\section{ME (\ref{ME-ciccarello}) as the continuous limit of \eq (\ref{drhon-gen})}

The derivation of ME (\ref{ME-ciccarello}) from \eq(\ref{drhon-gen}) is a slight generalisation of the analogous task carried out in \rref\cite{ciccarello} (see also \rref\cite{ciccarelloSM}).
When \eq(\ref{drhon-gen}) is divided by $\tau$ and using the limiting expressions discussed in the main text [see \eq(\ref{p-Gamma}) and related discussion], the terms on the right-hand side of \eq(\ref{drhon-gen}) in the continuous-time limit take the form
\begin{eqnarray}\label{fd2}
&&\frac{q \sum_{j=1}^{n\meno2}p^{(j-1)}\mathcal{E}_j\left[{\rho_{n\meno j}\meno\rho_{n\meno1\meno j}}\right]}{\tau}\!\simeq\!\Gamma\! \int_{0}^t \!dt' e^{-\Gamma t'}\mathcal{E}(t')\!\!\left[\frac{{\rm d}\rho(t\meno t')}{{\rm d} (t\meno t')}\right],\,\,\nonumber\\
&& \frac{qp^{n\meno1} \mathcal{E}_{n\meno1}}{\tau}\,[\rho_1]\!\simeq\!\Gamma e^{-\Gamma t}\mathcal{E}(t) [\rho_0]\,\,\nonumber\\
&&\frac{\Delta (p^{n\meno1} \Phi_n)}{\tau}\ug\frac{(p^{n\meno1} \Phi_n\meno p^{n\meno2} \Phi_{n\meno1})}{\tau}\nonumber\\
&&
\hspace{1cm}
\,\,\,\,\,\,\,\,\,\,\,\,\,\,\simeq\!\frac{e^{-\Gamma t}\Phi(t)\meno e^{-\Gamma (t- \tau)}\Phi(t-\tau)}{\tau}\!\simeq\! \frac{\rm d}{{\rm d} t}\!\left(e^{-\Gamma t }\Phi(t)\right).\nonumber
\end{eqnarray}
On the other hand, $\Delta\rho_n/\tau\!\rightarrow\!\dot\rho(t)$. 
By plugging all the above expressions into \eq(\ref{drhon-gen}), ME (\ref{ME-ciccarello}) is obtained.

\section{Proof of \eq(\ref{map-S})}

Let $\{|\nu\rangle_n\}$ be the set of eigenstates of state $\eta_n$, i.e., 
$$\eta_n=\sum_\nu p_{\nu} |\nu\rangle_n\langle \nu|\,,\label{etan}\,.$$
The set $\{|\nu\rangle_M\}$ is also a basis for the $M$'s Hilbert space (having the same dimension as $M$).
The swap unitary operator can then be expressed as
\begin{eqnarray}
\hat S_{Mn}=\sum_{\nu,\nu'}|\nu,\nu'\rangle_{M\!n}\langle \nu',\nu|\,.
\end{eqnarray}
Thereby, $\eta_n\hat S_{Mn}\ug\sum_{\nu,\nu'}p_{\nu'}|\nu\rangle_{M}\langle \nu'|\!\otimes\!|\nu'\rangle_{n}\langle \nu|$, which once plugged into \eq(\ref{map-S}) yields
\begin{eqnarray}
\mathcal S(\rho_{S\!M})&\ug&\Tr_{n}\left\{\hat S_{Mn}\rho_{S\!M}\,\eta_n\hat S_{Mn}\right\}\nonumber\\
&\ug&\sum_{\mu,\mu'}\!\sum_{\nu,\nu'}p_{\nu'}\!{_M}\!\langle \mu'|\rho_{S\!M}|\nu\rangle\!_M|\mu\rangle\!_M\!\langle \nu'|{\rm Tr}_n\left\{|\mu'\rangle\!_n\!\langle \nu|\right\}\!\delta_{\mu,\nu'}\nonumber\\
&\ug&\sum_{\mu,\nu}p_{\mu}\,{_M}\!\langle \nu|\rho_{S\!M}|\nu\rangle\!_M\,|\mu\rangle\!_M\!\langle \mu|\ug {\rm Tr}\{\rho_{S\!M}\}\,\eta_{M}\,.
\end{eqnarray}

\section{Induction proof of \eq(\ref{rhoS1-n})}\label{app-ind}

By construction, the $(n\piu1)$th-step state is related to the $n$th one as
\begin{equation}
\rho^{(n+1)}_{S\!M}\ug p~\U\!\left[\rho^{(n)}_{S\!M}\right]\piu q ~\U \mathcal S \left[\rho^{(n)}_{S\!M}\right]\,.\label{rhoS1-n+1}
\end{equation}
Our task is to show that if \eq(\ref{rhoS1-n}) holds then \eq(\ref{rhoS1-n+1}) can be arranged in the same form as \eq(\ref{rhoS1-n}) for $n\!\rightarrow\!(n\piu1)$.
Substituting \eq(\ref{rhoS1-n}) for $\rho^{(n)}_{S\!M}$ in the first term on the right-hand side of \eq(\ref{rhoS1-n+1}) yields
\begin{eqnarray}
\rho^{(n+1)}_{S\!M}&\ug&p\, \U\left\{\!\! q \!\sum_{j=1}^{n-1}p^{j-1}\,\U^j\mathcal S \left[\rho^{(n-j)}_{S\!M}\right]\!\!+\!p^{n-1}\U^n\left[\rho^{(0)}_{S\!M}\right]\!\right\}\nonumber\\
&&+ q ~\U\mathcal S \left[\rho^{(n)}_{S\!M}\right]\nonumber\\
&=& q \sum_{j=1}^{n-1}p^{j}~\U^{j+1}\mathcal S \left[\rho^{(n-j)}_{S\!M}\right]\!+\!p^{n}\U^{n+1}\left[\rho^{(0)}_{S\!M}\right]\nonumber\\
&&+ q ~\U\mathcal S \left[\rho^{(n)}_{S\!M}\right]\nonumber\,.
\end{eqnarray}
The last term on the right-hand side can be included in the sum over $j$ by making the index $j$ start from $j\ug0$. Carrying out next the index change $j\piu 1\!\rightarrow\! j$, we thus end up
with
\begin{eqnarray}
\rho^{(n+1)}_{S\!M}&\ug& q \sum_{j=1}^{n}p^{j-1}~\U^{j}\mathcal S\! \left[\rho^{(n+1-j)}_{S\!M}\right]\piu p^{n}\U^{n+1}(\rho^{(0)}_{S\!M})\nonumber\,,
\end{eqnarray}
which coincides with \eq(\ref{rhoS1-n}) for $n\!\rightarrow\!n\piu1$. This concludes our induction proof.


\begin{thebibliography}{99}
\bibitem{nota-closed} It is known that in such a case one can work out a closed ME for the atom with time-dependent rates \cite{petruccione}. These however depend on the time-evolved excitation probability amplitude of the atom, which in fact specifies the atom's full dynamical map \cite{petruccione}. In other words, this ME is closed but expressed in terms of its solution. Here, by ``closed ME" we mean one that does not depend on its solution.
\bibitem{nota-kossa} By comparing \eq(\ref{ffd}) and \eq(9) of \rref\cite{kossa}, we can identity the pair of ``legitimate" CP maps $N(t)=e^{-\Gamma t}\Phi(t)$, $Q(t)=\Gamma e^{-\Gamma t}\mathcal E(t)$ which, as required in \rref\cite{kossa}, indeed fulfil ${\rm Tr}\{Q(t)[\rho]+\partial _t {\rm Tr}\{N(t)[\rho]\}=0$ for any state $\rho$.
\end{thebibliography}
\end{document}